\documentstyle[aaspp4]{article}

\begin{document}
\title{Hard X-ray Variability in M82:Evidence for a Nascent AGN?}

\author{A. Ptak\altaffilmark{1} \& R. Griffiths}
\affil{Carnegie Mellon University, Dept. of Physics, Pittsburgh, PA 15213}
\altaffiltext{1}{email: ptak@astro.phys.cmu.edu}
\slugcomment{Submitted 12/08/98, Revised 2/18/99, 3/10/99, 3/22/99}
\abstract{
We report on the detection of hard (2-10 keV) X-ray
variability in the starburst
galaxy M82 over the course of 9 {\it ASCA} observations.
Long-term variability occurred on a time scale of days, with a
change in flux by a factor of up to $\sim 4$, corresponding to a point-source
luminosity of $L_{2-10\rm \ keV} \sim 6 \times 10^{40} \rm \ ergs \ s^{-1}$.
Short-term variability with an amplitude of $\sim 1.4$ 
on a timescale of hours was observed during the longest observation.
This demonstrates that a large fraction of the hard X-ray
emission of M82 (depending on the flux state) is from a compact
region and is probably due to an accreting source.
The 2-10 keV
luminosity of the source
is a lower limit to its Eddington luminosity, implying
a blackhole mass of at least $\sim 460M_{\odot}$, or a mass intermediate
to that of normal AGN and stellar-mass blackhole candidates.
}


\section{Introduction}
Hard X-rays (2 -- 10 keV) were marginally detected from M82 by {\it Uhuru}
(Forman et al. 1978), {\it Ariel-5}
(Cooke et al. 1978; McHardy et al. 1981), and
{\it HEAO-1} (Piccinotti et al. 1982),
making it one of the earliest extragalactic hard X-ray sources,
accurately positioned using the HEAO-A3 instrument  (Griffiths et
al. 1979).
These
detections were confirmed by the {\it Einstein MPC} (Watson, Stanger, \&
Griffiths 1984; Fabbiano 1988), {\it EXOSAT ME} (Schaaf et al.
1989) and {\it Ginga} (Ohashi et al. 1990).  Schaaf et al. speculated that
the hard X-ray flux from M82 was due to inverse-Compton scattering of infrared
photons off relativistic electrons (originally suggested by Hargrave
[1974]).  Ohashi et al. claimed that a bremsstrahlung fit was preferred over a
non-thermal power-law for the 2-20
keV {\it Ginga} spectrum, suggesting a thermal origin to the flux.  However,
the lack of a strong (equivalent width $\sim 1$ keV) Fe-K line detection
is problematic
for a thermal model, unless the abundances in the nucleus of M82 are
significantly subsolar (which would not be expected to be the case for a 
starburst galaxy).  A short {\it BBXRT} observation found that the 0.5-10.0
keV spectrum of M82 consists 
of at least two components, with the hard component described well by either
a power-law model or a thermal plasma model (Petre 1993).  M82 was originally
observed by {\it ASCA} in 1993 during
the Performance Verification (PV) phase of the mission.  These data confirmed
the {\it BBXRT} results 
and were analyzed by several groups
(Moran \& Lehnert 1997; Ptak et al. 1997; Tsuru et al. 1997).

Moran \&
Lehnert came to the conclusion that the hard flux originated in IC scattering
of infrared flux based on the good correlation of {\it ROSAT} ``hardness''
and infrared images.  Ptak et al. speculated that a variable X-ray binary
detected by {\it Einstein} (Watson, Stanger \& Griffiths 1984) and {\it ROSAT} 
(Collura et al. 1994) {\it could} account for a large fraction of the 2-10
keV flux from M82 {\it if} its X-ray spectrum was similar to the spectrum of
the hard component in M82.  Similarly, Tsuru et al. suggested that M82 has
varied significantly in the 2-10 keV bandpass, and accordingly the hard
X-ray emission is most likely due to an accreting binary or AGN.  However,
all fluxes prior to {\it ASCA} were from {\it non-imaging} detectors, and
accordingly any flux comparison is inherently suspect (i.e., due to the 
collimation of diffuse flux and spurious sources in the non-imaging detector's
field of view).  For example, M81, which is a variable 2-10 keV source,
lies $\sim 40'$ from M82.  Here we report on 2-10 keV
intensity and spectral variability detected in M82 from a series of monitoring
observations by {\it ASCA}.

\section{The {\it ASCA} Data}
The {\it ASCA} observations were performed during the period Mar. 6, 1996
through Nov. 11, 1996 at 9 intervals (see Table 1).
Briefly, {\it ASCA} (Tanaka et al. 1994) comprises two solid-state imaging spectrometers 
(SIS; hereafter S0 and S1) with an approximate bandpass of 0.4-10.0 
keV and two gas imaging spectrometers (GIS; hereafter G2 and 
G3) with an approximate bandpass of 0.8-10.0 keV. The SIS observations 
were done in 1-ccd mode\footnote{see
http://adfwww.gsfc.nasa.gov/asca/processing\_doc/GS/intro.html\#highlights},
and due to calibration problems below 0.6 keV
we only consider SIS data in the 0.6-10.0 keV bandpass (most of our
conclusions are based on data in the 2.0-10.0 keV bandpass). 
Because the 1993 observation showed that the 2-10 keV emission
from M82 is unresolved (Tsuru et al. 1997, Ptak et al. 1999), we chose source
regions appropriate for a point 
source, namely 6' for the GIS detectors and 4' for the SIS detectors.
The background 
was determined from the remaining counts in the CCDs beyond 
5' for the SIS, and from a 7-12' annulus for the GIS data. 
The spectra from each set of detectors were combined to yield a 
SIS and GIS spectrum for each observation, although we also performed
fits to each spectrum separately to check for anomalies.  Prior to spectral
fitting the data were binned to a minimum of 20 counts per bin to allow
the use of the $\chi^2$ statistic.

\begin{deluxetable}{ccclll}
\tablewidth{30pc}
\tablecaption{Observation Log}
\tablehead{
\colhead{ID} & \colhead{Date} & \colhead{Duration} & 
\colhead{Exposure} & \colhead{Count Rates\tablenotemark{a}}
& \colhead{Angle\tablenotemark{b}}
\nl
&  & \colhead{(ks)} & \colhead{(ks)} & \colhead{(cts/s)} & 
\colhead{(arcmin)}}
\startdata
1 & 3/23/96 & 20.0 & 12.9-13.6 & 0.89,0.73,0.47,0.60 & 5.7,7.8,7.5,5.1 \nl
2 & 4/15/96 & 18.2 & 6.9-8.7 & 1.25,1.07,0.75,0.95 & 5.8,8.2,7.6,5.3 \nl
3 & 4/21/96 & 24.1 & 12.4-12.9 & 1.15,0.94,0.66,0.84 & 5.8,8.1,7.6,5.2 \nl
4 & 4/24/96 & 64.3 & 29.3-33.3 & 1.15,0.92,0.66,0.82 & 5.8,8.3,7.7,5.5 \nl
5 & 5/13/96 & 18.6 & 8.1-9.4 & 0.78,0.64,0.41,0.51 & 6.2,8.6,8.0,5.7 \nl
6 & 5/5/96 & 19.3 & 11.5-14.4 & 0.91,0.76,0.49,0.60 & 6.2,8.6,8.0,5.7 \nl
7 & 10/14/96 & 25.5 & 7.6-8.6 & 0.59,0.51,0.30,0.36& 6.5,8.9,8.4,5.9 \nl
8 & 11/14/96 & 20.4 & 12.0-12.5 & 0.68,0.56,0.33,0.43 & 6.1,8.2,7.9,5.5 \nl
9 & 11/26/96 & 14.0 & 7.6-9.8 & 0.76,0.67,0.43,0.56& 5.9,7.7,7.6,4.7
\tablecomments{The count rates cited are for the full bandpass of each
detector ($\sim 0.6-10.0$ keV and $\sim 0.8-10.0$ keV for the SIS and GIS,
respectively).  The signal-to-noise ratio of the observations varied from
45-177, with a mean of 88.}
\tablenotetext{a}{Background-subtracted count rates for S0, S1, G2, G3.}
\tablenotetext{b}{Off-axis angle for S0, S1, G2, G3.}
\enddata
\end{deluxetable}

\section{Results}
\subsection{Long-term Variability}
A good fit to the M82 {\it ASCA} PV spectra 
required at least two components: a thermal plasma model dominating the soft
flux below 2 keV and a power-law or thermal bremsstrahlung model dominating
above 3 keV (Tsuru et al. 1997, Ptak et al. 1997, Moran \& Lehnert 1997).
Accordingly, we fit the two spectra from each observation with a thermal
plasma plus a 
power-law model, with only the power-law (dominating above 2-3 keV) results
being of interest in the 
present work.  Note that our motivation for fitting the spectra here is for
the purpose of determining fluxes rather than determining the precise values
of spectral parameters.  These fits were acceptable ($\chi^2_{\nu} =
0.95-1.10$).  
Figure 1 shows the 2-10 keV
long-term lightcurve based on (observed) fluxes
inferred from these fits.
\begin{figure}[htbn]
\epsscale{0.5}
\plotone{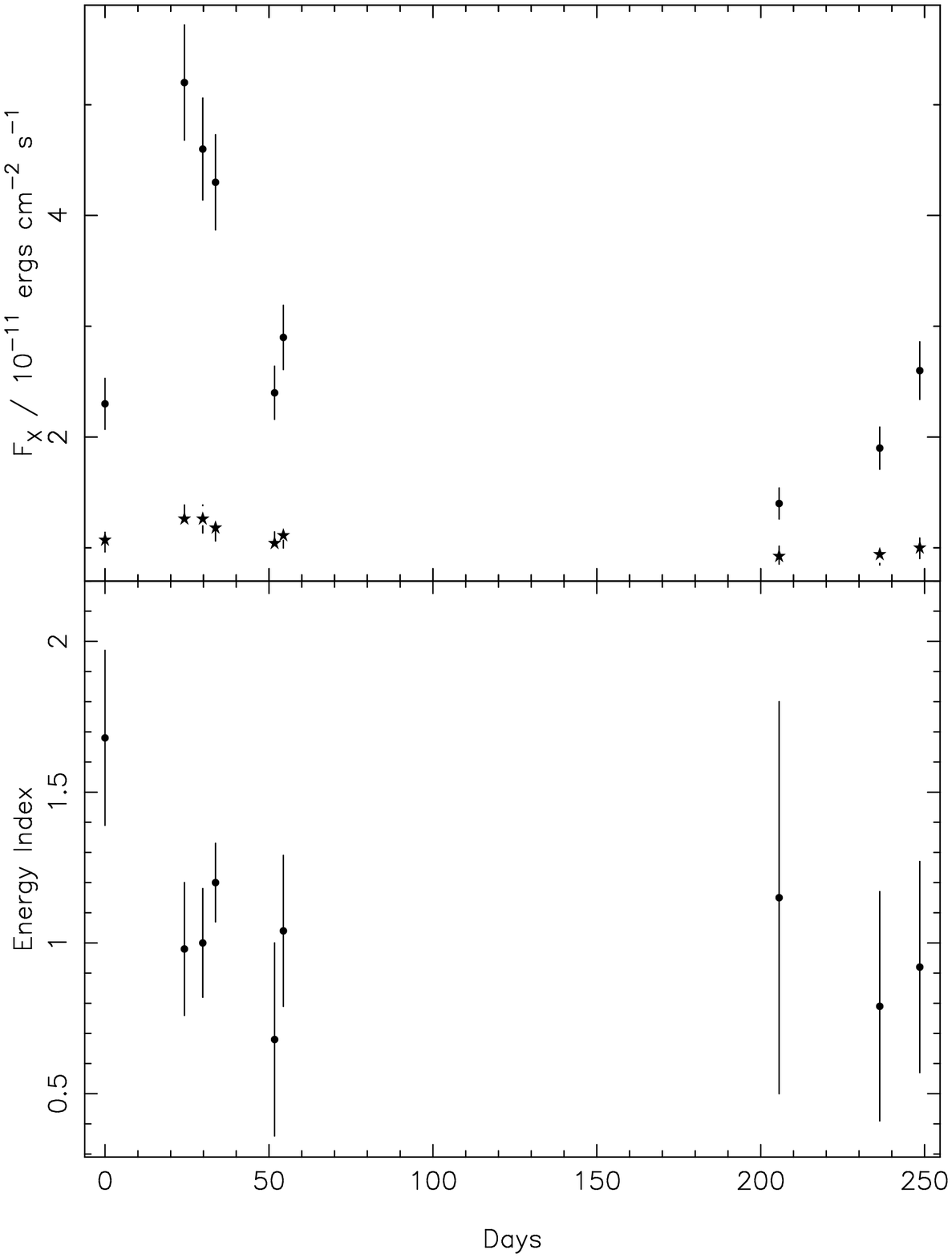}
\caption{\footnotesize
top) The 0.5-2.0 keV (filled stars) and 2-10 keV (filled circles)
long-term light curve from M82.  The fluxes are
not corrected for extinction and are inferred from a two-component fit to
the integrated spectrum from each observation.  The error bars are at a level
of $10\%$, which is a reasonable estimate of the relative uncertainty in flux
from observation to observation.  (bottom) The best-fitting power-law energy
indices resulting from 
the fits used to determine the fluxes.  The error bars show the 90\%
confidence intervals (see text).  The power-law dominates the 2-10 keV
bandpass.}
\end{figure}
We also fit the
data in the 3-10 keV bandpass only
with a simple power-law and found similar values for both fluxes and the
power-law slope, showing that the
soft component is not adversely affecting our results (in both sets of fits
the mean column density absorbing the power-law was on the order of $2 \times
10^{22} \rm \ cm^{-2}$, similar to the PV observation value of $1.9 \pm 1.0
\times 10^{22} \rm \ cm^{-2}$ given in Ptak et al. 1997).  Long-term
variability 
is evident, with a change in 2-10 keV flux from $\sim 1.4 \times 10^{-11}
\ \rm ergs \ cm^{-2} \ s^{-1}$ to $\sim 5.2 \times 10^{-11}
\ \rm ergs \ cm^{-2} \ s^{-1}$.
The soft flux from M82 is known to be extended, with some contribution
from point sources that are variable (c.f., Collura et al. 1994, Bregman
Schullman \& Tomisaka 1995). 
The 0.5-2.0 keV fluxes derived from the same fits (also shown in Figure 1)
remained within $\sim 15\%$
of the mean, demonstrating that the variability is occurring predominately 
above 2 keV.  The
stability of the soft flux also indicates that the observed variability is not
due to
instrumental effects (which would likely affect the entire bandpass in a
similar fashion) or the
explosion of a supernova which would likely result in a softer spectrum.
The mean 0.5-2.0 keV flux in these observations is $1.1 \times 10^{-11} \ \rm
ergs \ cm^{-2} \ s^{-1}$, comparable to the value cited in Ptak et al. (1997)
for the PV observation ($9.4 \times 10^{-12} \ \rm
ergs \ cm^{-2} \ s^{-1}$; the 2-10 keV PV flux is consistent with the
flux from the 8th observation given here).  Note that some of the variability
below 2 keV may be due to the same source(s) causing the hard variability, but
the precise quantification of this is uncertain since the level of variability
is on the order of systematic errors.
The power-law energy index
resulting from these fits is shown in the bottom panel of Figure 1.
The errors bars shown are the 90\% confidence
based on $\Delta\chi^2 = 17$ (11 interesting parameters), with similar results
being obtained for 
$\Delta\chi^2=4.6$ (2 interesting parameters) in the 3-10 keV fits.  The
typical $90\%$ confidence
on the energy index was $\sim 0.3$.  In most of the observations,
the energy index was consistent with the mean value of 1.0, however,
the energy index in the first observation was significantly steeper,
$\alpha = 1.68$ (1.42-1.97) (the PV observation power-law slope was
$0.76^{+0.22}_{-0.21}$).

\begin{figure}[htbn]
\epsscale{0.5}
\plotone{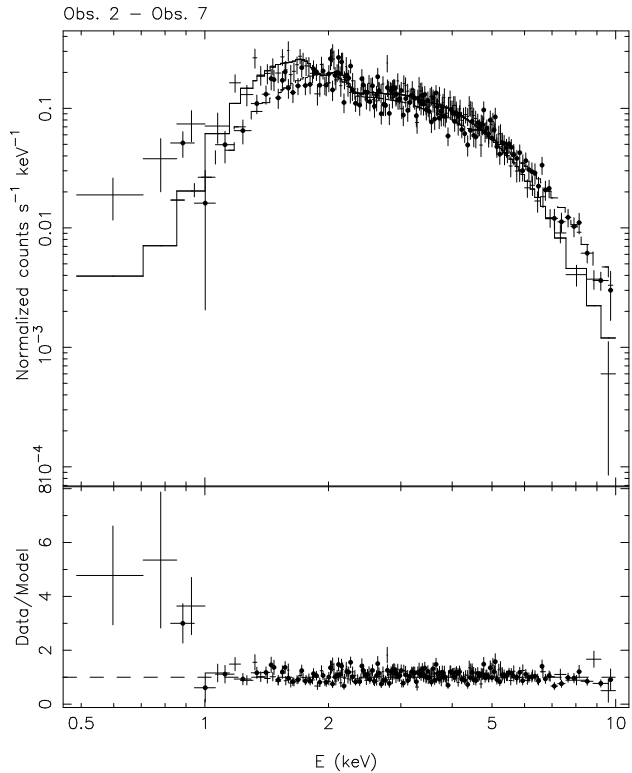}
\caption{\footnotesize
The resultant spectrum when the lowest flux spectrum is subtracted
from the highest flux spectrum.  The GIS data are marked with filled circles,
the SIS data are unmarked.  A power-law fit to the data is shown with a solid
line (SIS data) and a dotted line (GIS data).  The ratio of the data to the
model is shown in the bottom panel.}
\end{figure}
We also investigated the residual spectrum when the lowest flux state spectrum
is subtracted as background from the other spectra.  The motivation for
this was to isolate the spectrum of the variable source by subtracting the
(extended) soft flux, as well as any other sources
of hard X-ray flux (thermal emission, IC scattered photons, other
point sources, and background).  We subtracted the spectra prior to any binning
(or background subtraction). 
Figure 2 shows the results of a power-law fit to the highest
flux spectrum residuals. 
The power-law fit is excellent
in the 2-10 keV bandpass, as expected, but interestingly a soft ``tail'' is
also present, at a flux level (in this case) of $\sim 1 \times 10^{-12} \ \rm
ergs \ cm^{-2} \ s^{-1}$, or about $10\%$ of the total 0.5-2.0 keV flux from
M82, which is typical of the scatter in the 0.5-2.0 keV lightcurve shown in
Figure 1.  A detailed
discussion of these fits is beyond the scope of this paper, but we
note that the soft excess
must be treated cautiously since a 10\% variation could
be a systematic effect resulting from the subtraction of separate observations
(although note this component appears consistently among the SIS and GIS data
suggesting that this is not the case).

Since the pointing accuracy of {\it ROSAT} is significantly better than that
of {\it ASCA}, and the 0.5-2.0 keV flux observed by both satellites must have
a common origin, the position of the hard X-ray flux can best be
estimated by comparing the centroids of the hard and soft X-ray emission.
We found that the centroids
of the 0.5-2.0 keV and 3.0-10.0 keV flux are coincident to within $\sim 10''$
($\sim 200$ pc), indicating that the emission is nuclear, i.e., the 0.5-2.0 keV
emission is known to originate in the nucleus from {\it Einstein} and {\it
ROSAT} HRI observations (Watson, Stanger \&
Griffiths 1984; Bregman et al. 1995). We investigated the
radial profile distribution of the lowest flux state observation (ID 7 in
Table 1) and of the most variable observation (ID 4, see below) and found
that the 3-10 keV emission in both cases is unresolved.  A radial Gaussian fit
to the SIS0 radial profiles (see Ptak 1997, Ptak et al. 1999 for the method)
yielded an upper limit to the half-light radius of the emission on the order
of 30'' ($\sim 0.5$ kpc).  These extents are comparable to the spatial extent
found in Ptak (1997) and Ptak et al. (1999) for the {\it ASCA} PV M82 SIS
data.
This result
indicates that the source of variability is most likely a single nuclear
source rather than several X-ray binaries distributed over M82 (although a
small {\it nuclear} population of binaries cannot be excluded as a
possibility).

\subsection{Short-Term Variability}
\begin{figure}[htbn]
\epsscale{0.5}
\plotone{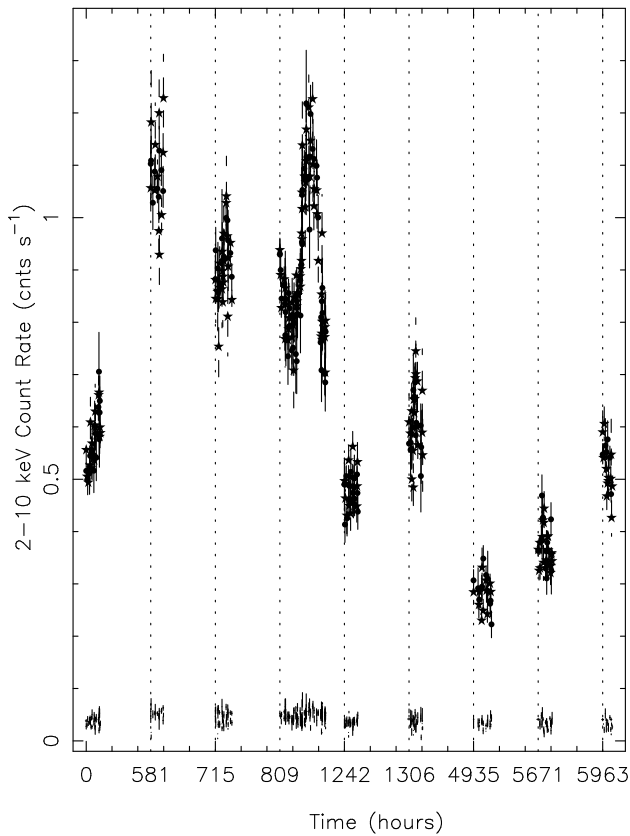}
\caption{\footnotesize
The short-term, 2-10 keV SIS (dashed lines marked with stars) and
GIS lightcurve (solid lines marked with filled circles) resulting from
each individual lightcurve concatenated onto a single plot.  The duration of
each observation (delimited by vertical dotted lines) is 25 hours, and X-axis
labels
give the start times of the observations.  The background
rate is also shown in the figure (at a mean rate of $0.031 \rm \ counts \
s^{-1}$, $\sim 5\%$ of the total, and $0.044 \rm \ counts \
s^{-1}$, $\sim 7\%$ of the total, for the SIS and GIS, respectively.).  Note
that given the {\it ASCA} PSF, several percent of the source flux is scattered
into the background regions used.  The size of each bin was $\sim 4000$ s.}
\end{figure}
Figure 3 shows the short-term lightcurves of each observation appended
into a single plot.
The lightcurve from the 4th observation is shown in 
Figure 4, where a large variation is evident.  The source varied by $\sim
20\%$ from the mean flux of the observation, or a peak-to-valley variation
of $\sim 1.4$.  This type of fluctuation was observed so clearly
in this observation only, with hints of variability (but typically at a lower
level) in the remaining observations.  A sine function fit to the lightcurve
in Figure 4 results in a period of $\sim 15$ hours.  We ``detrended'' the
long-term light-curve by subtracting the mean count rate of each individual
observation, and found that trend observed in Figure 4 does not extend to the
other lightcurves, i.e., this ``QPO'' is not consistent with an underlying
periodicity superimposed on long-term variations.  An alternative possibility
is that this is a flare with a duration 
of $\sim 15$ hours that caused an increase in the flux by $\sim 40\%$.
Note, however, that
the durations of the 
other observations were typically only 
5 hours, so strong conclusions cannot be drawn concerning the nature
of the short-term variability until a longer, contiguous
observation of M82 is made in the 2-10 keV bandpass.  
\begin{figure}[h]
\epsscale{0.5}
\plotone{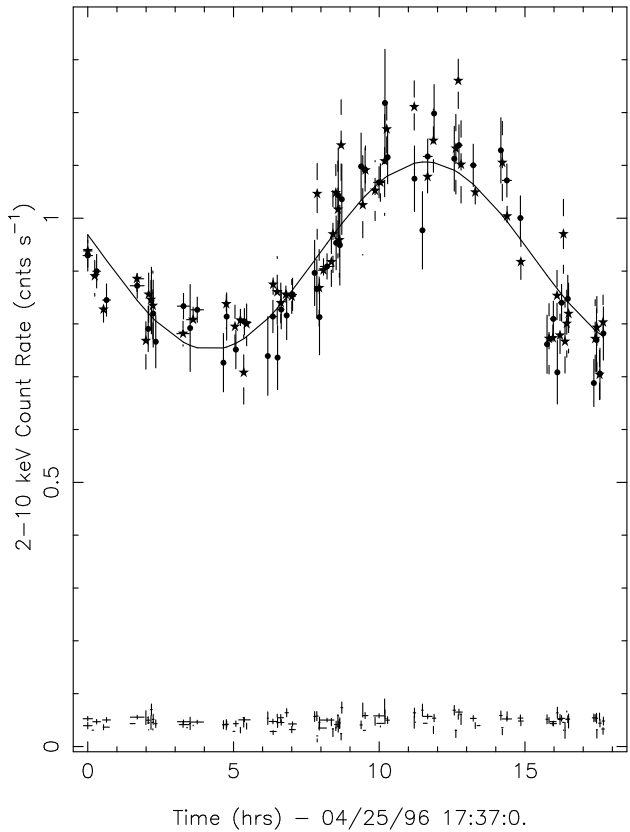}
\caption{\footnotesize
The short-term, 2-10 keV SIS and GIS lightcurves (marked as in
Figure 3) from the 4th observation
(the observation with the longest duration) is shown.  The curve shows a
sinusoidal fit to the GIS data (similar results are obtained from the SIS
data), with a best-fit period of 14.7 hours 
($\chi^2$/dof = $100.2/50$).}
\end{figure}

\section{Discussion}
We have presented data showing that a variable source of 2-10 keV flux is 
present in M82, located within $\sim 10''$ (170 pc) of the nucleus.
This source may be associated with the variable
X-ray source discussed in Collura et al., although higher angular resolution
is necessary to assess this possibility (with {\it ASCA} we can only note
that the hard and soft flux are coincident to within $\sim 10''$, which the
Collura et al. source, which can dominate the soft flux, is within $\sim 5''$
of the nucleus of M82).  
Here for the first time we
show that a variable, compact source is not only producing hard emission, but
can also {\it dominate} the hard flux from M82.
There is a suggestion of periodic or semi-periodic variability,
similar to that observed in some galactic X-ray binaries (c.f., Verbunt
1993).

The marginal detection
of Fe-K$\alpha$ emission at $\sim 6.7$ keV
in the 1993 {\it ASCA} observation (with M82 at a 2-10 keV flux of $\sim 2.0
\times 10^{-11} \ \rm ergs \ cm^{-2} \ s^{-1}$)  implies that at least some
of the hard component is thermal (Ptak et al. 1997, see also the {\it
BeppoSAX} Fe-K detection given in Cappi et al. 1999a).
The source also appears to exhibit spectral variability,
although predominately just between the first and remaining observations.
No correlation between spectral slope and flux is observed, although the
statistics are limited.

Assuming a distance to M82 of 3.6 Mpc (Freedman et al.
1994), the difference between the highest and lowest 2-10 keV fluxes observed
corresponds to a luminosity of $5.9 \times 10^{40} \ \rm ergs \ s^{-1}$.  For
comparison, the brightest nearby blackhole candidate (BHC) is SMC X-1 with a
2-10 keV luminosity of $6 \times 10^{38} \ \rm ergs \ s^{-1}$ (Verbunt 1993).
The timescale of this variability requires the source of this hard X-ray flux
to be compact, and it is most likely an accreting system with a
lower-limit on the blackhole mass of $\sim 460 M_{\odot}$ unless the high flux
rate is
exceeding the Eddington limit for the system, the emission is anisotropic,
and/or the relativistic motions are significant.
Given the high luminosity of this source (i.e., in a galaxy not known to be
harboring an AGN), these caveats should not be taken lightly.
As discussed in Komossa \& Bade (1999), other exotic possibilities for
luminous variable X-ray sources such as this include
a radio supernova, the tidal disruption of a star by a massive blackhole,
and a gamma ray burst.  However, the lack of a monotonic decline in the
lightcurve is problematic for these possibilities\footnote{
Komossa \& Bade favor variability in the absorption (in either column
density or ionization state) of an AGN as the explanation for the observed
lightcurves of NGC 5905 and IC3599, but in our case the available of {\it
ASCA}
data above 2 keV rules out this possibility.}.
Taken at face value, this mass estimate
is intermediate between the $10^{6-9} M_{\odot}$ systems
found in active galactic nuclei (AGN) and stellar-mass blackhole systems, and
this source may represent either an existing low-luminosity AGN
(with a mass on the order of $10^{6}M_{\odot}$ and a very low accretion
rate), or a ``nascent'' AGN.  By the latter possibility we mean a
blackhole 
system that is gradual growing in mass to eventually become a massive AGN.  The
supernova rate in M82 is $\sim 0.1\rm \ yr^{-1}$ (Van Buren \&
Greenhouse 1994), and over
a typical starburst lifetime of $10^{7}$ years, of order $10^6$ solar masses
of compact matter (i.e., neutron stars and blackholes) should be produced.
We are therefore speculating that on the order of at least
$\sim 0.05\%$ of this matter may have coalesced in the nucleus.

It would be of interest to distinguish between the AGN and massive BHC
scenarios.  Note that the lightcurve of this source does not show the
exponential decay characteristic of blackhole novae (c.f., Figure 3 in 
Tanaka \& Shibazaki 1996), although a more systematic monitoring of M82 would
be necessary to compare the temporal properties of this source to Galactic
BHC.
If the accretion disk or region in this source is
optically-thick, 
then a $M > 10^7 M_\odot$ blackhole accretion disk would emit a blackbody
spectrum with a temperature less than $10^6$ K, while in contrast accretion
disk temperatures for a blackhole candidate  are typically on the order of
$10^{6-7}$ K (c.f., Frank et al. 1992).  Therefore, the {\it unambiguous}
detection of a soft ``excess'' that can be attributed to this source alone (as
should be possible with the arcsecond resolution of {\it AXAF}) would yield an
important clue.  For example,
Colbert \& Mushotzky (1999) have analyzed the X-ray data
available for a sample of nearby normal galaxies and found that bright,
extranuclear point-sources are common.  The soft spectra of these sources were
fit well with disk models suggesting blackhole masses in the
range of $10^{2-4} M\odot$, consistent with the source in M82 (which may
not have settled into the dynamical center of M82 yet).  
On the other hand, if the accretion mode is
advection-dominated, then there should be no soft excess (note that in this
case the relative steepness of the 2-10 keV emission implies that the
accretion rate must exceed $\sim 10^{-3}$ in Eddington units, c.f. Ptak et
al. 1998).  The unambiguous association between this source and counterparts in
other wavebands (particularly in the radio where AGN candidates in M82 have
been proposed; see Seaquist, Frayer \& Frail 1997 and references therein, but
also see Allen \& Kronberg 1998) 
would go long way towards resolving this issue.

 
\acknowledgements
This paper made extensive use of the NASA's High Energy Astrophysics Science
Archive Research Center and Astrophysics
Data System Abstract Service databases, and the NASA/IPAC Extragalactic
Database.  We would also like to thank the referee for useful comments.

\end{document}